\documentclass{ws-rv9x6}

\begin{document}

\setcounter{chapter}{0}

\chapter{SUPPLY AND PRODUCTION
NETWORKS: FROM THE BULLWHIP EFFECT TO
BUSINESS CYCLES}

\markboth{Dirk Helbing and Stefan L\"ammer}{Supply Networks: From the
Bullwhip Effect to Business Cycles}

\author{Dirk Helbing and Stefan L\"ammer}

\address{Institute for Transport \& Economics,
Dresden University of Technology\\
Andreas-Schubert-Str. 23, D-01062 Dresden, Germany\\
E-mail: helbing@trafficforum.org, traffic@stefanlaemmer.de}

\begin{abstract}
Network theory is rapidly changing our understanding of complex systems,
but the relevance of topological features for the dynamic
behavior of metabolic networks, food webs, production systems, information networks,
or cascade failures of power grids remains to be explored. Based on a simple model of
supply networks, we offer an interpretation of instabilities and oscillations observed in biological,
ecological, economic, and engineering systems.
\par
We find that most supply networks display damped
oscillations, even when their units~-- and linear chains of these units~-- behave in a non-oscillatory
way. Moreover, networks of damped oscillators tend to produce growing oscillations. This surprising
behavior offers, for example, a new interpretation of business cycles and of oscillating or pulsating
processes. The network structure of material flows itself turns out to be a source of instability, and
cyclical variations are an inherent feature of decentralized adjustments.
In particular, we show how to treat production and supply networks as transport problems
governed by balance equations and equations for the adaptation of production speeds.

The stability and dynamic behavior of supply networks is investigated for
different topologies, including sequential supply chains, ``supply circles'', ``supply ladders'',
and ``supply hierarchies''. Moreover,
analytical conditions for absolute and convective instabilities are derived. The empirically
observed bullwhip effect in supply chains is explained as a form of
convective instability based on resonance effects. An application of
this theory to the optimization of production networks has large
optimization potentials.
\end{abstract}

\section{Introduction}

Supply chain management is a major subject in economics, as it
significantly influences the efficiency of production processes.
Many related studies focus on subjects such as optimum buffer
sizes and stock levels.\cite{QSCM,Zipkin,Nahmias,FactPhys} However, the optimal
structure of supply and distribution networks is also an important
topic,\cite{mitWitt,seba} which connects this scientific
field with the statistical physics of
networks.\cite{nets1,nets2,nets3,nets4,nets5,nets6,nets7,Dorogovtsev}
Problems like the adaptivity
and robustness of supply networks have not yet been well covered
and are certainly an important field for future research.
\par
In this chapter, we will focus on the dynamical properties and
linear stability of supply networks in dependence of the network
topology or, in other words, the supply matrix (input matrix).
Presently, there are only a few results on this subject, since the
response of supply networks to varying demand is not a trivial
problem, as we will see. Some ``fluid models'' have, however, been
proposed to study this subject. Daganzo,\cite{Dag} for example, applies the
method of cumulative counts\cite{Newell} and a variant of the
Lighthill-Whitham traffic model to study the so-called bullwhip
effect.\cite{Forrester,Levi,Cohen,Kahn,Lee1,Lee,Metters,Dej02,Dej03,Disney,Hoberg1,Hoberg2}
This effect has, for example, been reported for
beer distribution,\cite{beer1,beer2} but similar dynamical effects
are also known for other distribution or transportation
chains. It describes increasing oscillations in the delivery rates and
stock levels from one supplier to the next upstream supplier
delivering to him.

Similar models are studied by Armbruster  {\em
et al.}\cite{Arm,Arm1,Arm2,Arm3} Most
publications, however, do not investigate the impact of the
network topology, but focus on sequential supply chains or
re-entrant problems.\cite{Arm3,TGF03} An exception is the transfer function
approach by Dejonckheere {\em et al.}\cite{Dej02} as well as Disney and Towill.\cite{Disney}
Ponzi, Yasutomi, and Kaneko\cite{Ponzi} have coupled
a supply chain model with a model of price dynamics, while
Witt and Sun\cite{Witt} have suggested to model business cycles analogously to
stop-and-go traffic. Similar suggestions have been made by Daganzo\cite{Dag}
and Helbing.\cite{NJP} We should also note the relationship with driven many-particle
models.\cite{inRadons} This is the basis of event-driven simulations of production
systems, which are often used in logistics software.
\par
In a previous publication, one of the authors has proposed a rather general model
for supply networks, and has connected it to queueing theory and
macroeconomics.\cite{NJP} That paper also presents numerical
studies of the bullwhip effect in sequential supply
chains and the effect of heterogeneous production units on the
dynamics of supply networks. Moreover, the dependence of the
maximum oscillation amplitude on the model parameters has been
numerically studied, in particular the phase transition from
stable to bullwhip behavior. The underlying simulation models are
of non-linear nature,\cite{Radons} so that phenomena such as wave selection and
synchronization have been observed.\cite{NJP} Finally, a large variety of
management strategies including the effect of forcasting inventories
has been studied together with Nagatani\cite{NJP,stabilization}. The latter
has also investigated the dynamical effect of multiple production lines.\cite{Nagatani}
\par
In the following review, we will mainly focus on a linearized model of supply networks, which
could be viewed as a dynamical version\cite{Jorgenson} of Leontief's classical
input-output model.\cite{Leontief} It is expected to be valid
close to the stationary state of the supply system or, in other
words, close to the (commonly assumed) economic equilibrium. This
model allows to study many properties of supply networks in an
analytical way, in particular the effects of delayed adaptation of
production rates or of a forecasting of inventories. In this way,
we will be able to understand many numerically and empirically
observed effects, such as resonance effects and the stability
properties of supply networks in dependence of the network
topology.
\par
Our review is structured as follows: Section~\ref{Model} introduces
our model of supply networks and linearizes it around the
stationary state. Section~\ref{Mod} discusses the
bullwhip effect for sequential supply chains. Surprisingly,
the amplification of amplitudes does not require unstable
eigenvalues, but it is based on a resonance effect. Sec.~\ref{Sol}
presents general methods of solution for arbitrary topologies of
supply networks. It reveals some useful properties of the
eigenvalues characterizing the stability of supply systems.
Interestingly enough, it turns out that all supply networks can be
mapped on formulas for linear (serial) supply chains.
Sec.~\ref{examples} will investigate, how the dynamic behavior of
supply networks depends on their network structure, while
Sec.~\ref{businesscycles} gives an interpretation of business
cycles based on material flow networks. Sec.~\ref{Sum} finally
summarizes our results and points to some future research
directions.

\section{Input-Output Model of Supply Networks} \label{Model}

Our  production model assumes $u$ production units or suppliers $j\in \{1,\dots,u\}$ which deliver
$d_{ij}$ products of kind $i\in \{1,\dots,p\}$ per production cycle to other suppliers and consume
$c_{kj}$ goods of kind $k$ per production cycle. The coefficients $c_{kj}$ and $d_{ij}$ are
determined by the respective production process, and the number of production cycles per
unit time (e.g. per day) is given by the production speed $Q_j(t)$. That is, supplier $j$ requires an average time
interval of $1/Q_j(t)$ to produce and deliver $d_{ij}$ units of good $i$. The temporal change in the number
$N_i(t)$ of goods of kind $i$ available in the system
is given by the difference between the inflow
\begin{equation}
Q_i^{\rm in}(t) = \sum_{j=0}^{u} d_{ij} Q_j(t)
\end{equation}
and the outflow
\begin{equation}
 Q_i^{\rm out}(t) = \sum_{i=1}^{u+1} c_{ij} Q_j(t) \, .
\end{equation}
In other words, it is determined by the overall production rates $d_{ij}Q_j(t)$ of all suppliers $j$ minus their
overall consumption rates $c_{ij}Q_j(t)$:
\begin{equation}
 \frac{dN_i}{dt} = Q_i^{\rm in}(t) - Q_i^{\rm out}(t)
 = \underbrace{\sum_{j=1}^u d_{ij} Q_j(t)}_{\rm supply}
 - \underbrace{\left[ \sum_{j=1}^u c_{ij} Q_j(t) - Y_i(t) \right]}_{\rm demand} \, .
\label{conserv}
\end{equation}
In this dynamic variant of Leontief's classical input-output model,\cite{Leontief} the quantity
\begin{equation}
 Y_i(t) = \underbrace{c_{i,u+1}Q_{u+1}(t)}_{\rm consumption\ and\ losses} \quad
 - \underbrace{d_{i0} Q_0(t)}_{\rm inflow\ of\ resources}
\end{equation}
comprises the consumption rate of goods $i$, losses, and waste (the ``export'' of material),
minus the inflows into the considered system (the ``imports''). In the following, we will assume that
the quantities are measured in a way that $0 \le c_{ij}, d_{ij} \le 1$  (for $1 \le i \le p$, $1 \le j \le u$)
and that the ``normalization conditions''
\begin{equation}
  d_{i0} = 1 - \sum_{j=1}^u d_{ij}  \ge 0 \, , \qquad c_{i,u+1} = 1 - \sum_{j=1}^u c_{ij} \ge 0
\end{equation}
are fulfilled. Equations (\ref{conserv}) can then be interpreted as conservation equations for the flows of goods.

\subsection{Adaptation of Production Speeds}

In addition, we have formulated an equation for the production
or delivery rate $Q_j(t)$. Changes in the
consumption rate $Y_i(t)$ sooner or later require an adaptation of $Q_j(t)$. For
the temporal change $dQ_j/dt$ of the delivery rate we will assume:
\begin{equation}
 \frac{dQ_j}{dt} = F_j(\{N_i(t)\},\{dN_i/dt\},\{Q_l(t)\}) \, .
\label{stra}
\end{equation}
Herein, the curly brackets indicate that the so-called management or control function
$F_j(\dots)$ may depend on all inventories $N_i(t)$ with $i\in \{1,\dots,p\}$, their
derivatives $dN_i/dt$, and/or all production speeds $Q_l(t)$ with $l \in \{1,\dots,u\}$.
Some reasonable specifications will be discussed later on.

\subsection{Modelling Sequential Supply Chains} \label{Mod}

\begin{figure}[htbp]
\begin{center}
\end{center}
\vspace*{8pt} \caption[]{Illustration of the linear supply chain
treated in this chapter, including the key variables of the model.
Circles represent different suppliers $i$, $N_i$ their respective
stock levels, and $Q_i$ the delivery rate to supplier $i$ or the
production speed of this supplier. $i=0$ corresponds to the
resource sector generating the basic products and $i=u+1$ to the
consumers.}
\label{sequential}
\end{figure}
For simplicity, let us first investigate a model of sequential
supply chains (see Fig.~\ref{sequential}), which corresponds to
$d_{ij} = \delta_{ij}$ and $c_{ij} = \delta_{i+1,j}$. In other
words, the products $i$ are directly associated with producers $j$
and we have an input matrix of the form
\begin{equation}
\mathbf{C} = \left(
\begin{array}{ccccc}
0 & 1 & 0 & 0 & 0 \\
0 & 0 & 1 & 0 & 0 \\
0 & 0 & 0 & 1 & 0 \\
0 & 0 & 0 & 0 & 1 \\
0 & 0 & 0 & 0 & 0 \\
\end{array}
\right) \, .
\label{serial}
\end{equation}
This implies $Q_i^{\rm in}(t)
= Q_{i}(t)=Q_{i-1}^{\rm out}(t)$ and $Q_i^{\rm out}(t) = Q_{i+1}(t) = Q_{i+1}^{\rm in}(t)$, so that
the inventory of goods of kind $i$ changes in time $t$ according to
\begin{equation}
\frac{dN_i }{dt} = Q_i^{\rm in}(t) - Q_i^{\rm out}(t) = Q_i (t) - Q_{i + 1} (t) \, .
\label{eq3}
\end{equation}
The assumed model consists of a series of $u$ suppliers $i$, which
receive products of kind $i-1$ from the next ``upstream'' supplier $i-1$
and generate products of kind $i$
for the next ``downstream'' supplier $i+1$ at a rate $Q_i(t)$.\cite{Lefeber,Arm}
The final products are delivered at the rate $Q_u(t)$ and removed from the system with the
consumption rate $Y_u(t) = Q_{u+1}(t)$.
\par
In the following, we give some possible specifications
of equation (\ref{stra}) for the production rates:
\begin{itemize}
\item The delivery rate $Q_i(t)$ may, for example, be adapted to a certain desired rate
$\hat{W}_i\left(N_i,dN_i/dt\right)$ according to the equation
\begin{equation}
 \frac{dQ_i}{dt} = \frac{1}{T_i} \left[ \hat{W}_i\left(N_i,\frac{dN_i}{dt}\right)
 - Q_i(t) \right] \, ,
\label{st1}
\end{equation}
where $T_i$ denotes the relaxation time.
\item A special case of this is the equation
\begin{eqnarray}
 \frac{dQ_i}{dt} &=& \frac{1}{T_i} \bigg[ W_i\Big(N_i(t)\Big)
+ \bigg( \underbrace{\frac{dW_i(N_i)}{dN_i} \; \Delta t}_{= \beta_i}
\bigg) \frac{dN_i}{dt} - Q_i(t) \bigg] \nonumber \\
&\approx & \frac{1}{T_i}
\bigg[ W_i\Big( N_i(t+\Delta t) \Big) - Q_i(t) \bigg] \, .
\label{mind}
\end{eqnarray}
Herein, the desired production speed $W_i(N_i)$ is monotonously falling with increasing
inventories $N_i$, which are forecasted over a time period $\Delta t$.
\item A further specification is given by the equation
\begin{equation}
 \frac{dQ_i}{dt} = \frac{1}{T_i} \left\{ \frac{N_i^0 - N_i(t)}{\tau_i}
 - \beta_i \frac{dN_i}{dt} + \epsilon_i[Q_i^0 - Q_i(t)] \right\} \, ,
\end{equation}
which assumes that the production rate is controlled in a way that
tries to reach some optimal inventory $N_i^0$ and production rate $Q_i^0$, and
attempts to miminize changes $dN_i/dt$ in the inventory to
reach a constant work in progress (CONWIP strategy). Note that at least
one of the parameters $T_i$, $\tau_i$, $\beta_i$ and $\epsilon_i$ could be dropped.
\item In other cases, it can be reasonable to work with a model focussing on
relative changes in the variables:
\begin{equation}
 \frac{dQ_i/dt}{Q_i} = \hat{\nu}_i \left( \frac{N_i^0}{N_i(t)} - 1 \right)
 - \hat{\mu}_i \frac{dN_i/dt}{N_i(t)} +\hat{\epsilon}_i \left( \frac{Q_i^0}{Q_i(t)} - 1 \right) \, .
\label{st2}
\end{equation}
This model assumes large production rates when the inventory is
low and prevents that $Q_i(t)$ can fall below zero. Apart from
this, $N_i(t)$ and $Q_i(t)$ are adjusted to some values $N_i^0$
and $Q_i^0$, which are desireable from a production perspective
(i.e. in order to cope with the breakdown of machines, variations in
the consumption rate, etc.) Moreover, the control strategy
(\ref{st2}) counteracts temporal changes $dN_i/dt$ in the
inventory. For comparison with the previous strategies, we set
$T_i = 1/Q_i^0$, $\tau_i = N_i^0/\hat{\nu}_i$, $\beta_i =
\hat{\mu}_i/N_i^0$, and $\epsilon_i = \hat{\epsilon}_i T_i$.
\end{itemize}
The above control strategies appear to be
appropriate to keep the inventories $N_i(t)$ stationary, to
maintain a certain optimal inventory $N_i^0$ (in order to cope
with stochastic variations due to
machine breakdowns etc.), and to operate with the equilibrium
production rates $Q_i^0 = Y_u^0$,
where $Y_u^0$ denotes the average consumption rate. However,
the consumption rate is typically subject to perturbations, which may
cause a bullwhip effect, i.e.
growing variations in the stock levels and deliveries of upstream suppliers
(see Sec.~\ref{Resonance}).

\subsection{More Detailed Derivation of the Production Dynamics (with Dieter Armbruster)}

Let us assume a sequential supply chain and let $Q_i(t)$ be the influx in the last time period $T =
24$ hours from the supplier $i-1$ to producer $i$. This influx is equal to the release
$R_{i-1}(t)$ over the last 24 hours:
\begin{equation}
  Q_i(t) = R_{i-1}(t) \, .
\end{equation}
Based on the current state (the inventory and fluxes at time $t$ and before), a price
$P_i(t)$ is set according to some pricing policy
and instantaneously communicated downstream to the customer $i+1$. Then,
based on price, fluxes, and inventory, the order quantity $D_{i+1}(t)$ is decided at the end of day $t$
according to some order policy $W_{i+1}$:
\begin{equation}
 D_{i+1}(t) =  W_{i+1}\Big(N_{i+1}(t),P_i(t),\dots\Big) \, .
\end{equation}
It determines the upstream release $R_i(t+T)$ on the next day $t+T = t + 1$:
\begin{equation}
 R_i(t+T) = R_i(t+1) = D_{i+1}(t) \, .
\end{equation}
Altogether, this implies
\begin{equation}
 Q_i(t+T) = Q_i(t+1) = R_{i-1}(t+1) = D_i(t) = W_i\Big(N_{i}(t),\dots\Big)
\end{equation}
and the following equation for the change in the production rate:
\begin{equation}
 \frac{dQ_i}{dt} \approx \frac{\Delta Q_i(t)}{T} = \frac{1}{T} [ Q_i(t+T) - Q_i(t) ]
 = \frac{1}{T} \Big[ W_i\Big(N_{i}(t),\dots\Big) - Q_i(t) \Big] \, .
\end{equation}
This is consistent with Eq.~(\ref{st1}).
Moreover, we obtain the usual balance equation for the change of inventory in time:
\begin{eqnarray}
 \frac{dN_i}{dt} &\approx & \frac{\Delta N_i}{T} = \frac{N_i(t+T) - N_i(t)}{T} = N_i(t+1) - N_i(t) =
 R_{i-1}(t) - R_i(t) \nonumber \\
 &=& Q_i(t) - Q_{i+1}(t) \, .
\end{eqnarray}

\subsection{Dynamic Solution and Resonance Effects (with Tadeusz P\l atkowski)}\label{Resonance}

Let us now calculate the dynamic solution of the sequential supply
chain model for cases where the values $N_i^0$
and $Q_i^0$ correspond to the stationary state of the production
system.\cite{seba,inRadons} Then, the linearized model equations for the control
approaches (\ref{st1}) to (\ref{st2}) are exactly the same.
Representing the deviation of the inventory from the stationary one
by $n_i(t) = N_i(t) - N_i^0$ and the deviation of the delivery
rate by $q_i(t) = Q_i(t) - Q_i^0$, they read
\begin{equation}
 \frac{dq_i}{dt} = \frac{1}{T_i} \left[ - \frac{n_i(t)}{\tau_i}
 - \beta_i \frac{dn_i}{dt} - \epsilon_i q_i(t) \right] \, .
\label{lin1}
\end{equation}
Moreover, the linearized equations for the inventories are given by
\begin{equation}
 \frac{dn_i }{dt} = q_i (t) - q_{i + 1} (t) \, .
\label{lin2}
\end{equation}
Deriving Eq.~(\ref{lin1}) with respect to $t$ and inserting Eq.~(\ref{lin2}) results in the
following set of second-order differential equations:
\begin{equation}
\frac{d^{2}q_i}{dt^{2}}+\underbrace{\frac{(\beta_i+\epsilon_i)}{T_i}}_{=2\gamma_i}
\frac{dq_i}{dt}+\underbrace{\frac{1}{T_i\tau_i}}_{=\omega_i{}^{2}}
q_i(t) =\underbrace{\frac{1}{T_i} \left[ \frac{q_{i+1}(t)}{\tau_i}
+ \beta_i \frac{dq_{i+1}}{dt} \right]}_{= f_i(t)} \, .
\label{set}
\end{equation}
This corresponds to the differential equation of a damped harmonic oscillator
with damping constant $\gamma$, eigenfrequency $\omega$, and driving term $f_i(t)$.
The eigenvalues of these equations are
\begin{equation}
\lambda_{1,2}=-\gamma_i \pm\sqrt{\gamma_i{}^{2}-\omega_i{}^{2}}
= - \frac{1}{2T_i} \Bigg[ (\beta_i+\epsilon_i) \mp \sqrt{(\beta_i+\epsilon_i)^2 - 4T_i/\tau_i} \Bigg] \, .
\end{equation}
For $(\beta_i+\epsilon_i) > 0$ their real parts are always negative, corresponding to a stable behavior
in time. Nevertheless, we will find a {\em convective} instability, i.e. the oscillation amplitude
can grow from one supplier to the next one upstream.
\par
Assuming periodic oscillations of the form $f_{i}(t)=f_{i}^{0}\cos(\alpha t)$,
the general solution of Eq. (\ref{set}) is of the form
\begin{equation}
q_{i}(t)=f_{i}^{0}F_i\cos(\alpha t +\varphi_i) + D_i^0 \mbox{e}^{-\gamma_i t}
\cos(\Omega_i t + \theta_i) \, ,
\end{equation}
where the parameters $D_i^0$ and $\theta_i$ depend on the initial conditions.
The other parameters are given by
\begin{equation}
 \tan \varphi_i =\frac{2\gamma_i\alpha}{\alpha{}^{2}-\omega_i{}^{2}}
 = \frac{\alpha(\beta_i+\epsilon_i)}{\alpha{}^2 T_i - 1/\tau_i} \, ,
\end{equation}
\begin{equation}
 \Omega_i = \sqrt{\omega_i{}^2 - \gamma_i{}^2} = \frac{1}{T_i}\sqrt{1/\tau_i{}^2 - (\beta_i + \epsilon_i)^2/4} \, ,
\end{equation}
and
\begin{equation}
 F_i = \frac{1}{\sqrt{(\alpha{}^{2}-\omega_i{}^{2})^{2}+4\gamma_i{}^{2}\alpha{}^{2}}}
 = \frac{T_i}{\sqrt{(\alpha{}^{2}T_i-1/\tau_i)^2 + \alpha{}^2(\beta_i+\epsilon_i)^2}} \, .
\end{equation}
The dependence on the eigenfrequency $\omega_i$ is important for understanding the
occuring resonance effect, which is particularly likely to appear, if the
oscillation frequency $\alpha$ of the consumption rate is close to one of the
resonance frequencies $\omega_i$.
After a transient time much longer than $1/\gamma_i$ we find
\begin{equation}
q_{i}(t)=f_{i}^{0}F_i\cos(\alpha t +\varphi_i) \, .
\label{in}
\end{equation}
Equations (\ref{set}) and (\ref{in}) imply
\begin{equation}\label{peitscheneffekt1}
f_{i-1}(t)=\frac{1}{T_i}\left[\frac{q_{i}(t)}{\tau_i} + \beta_i \frac{dq_{i}}{dt}\right]
= f_{i-1}^{0}\cos(\alpha t + \varphi_i + \delta_i)
\end{equation}
with
\begin{equation}
\tan \delta_i =\alpha \beta_i \tau_i
\qquad \mbox{and} \qquad
f_{i-1}^{0}= f_{i}^{0} \frac{F_i}{T_i} \sqrt{(1/\tau_i)^2+(\alpha\beta_i)^{2}} \, .
\end{equation}
Therefore, the set of equations (\ref{set}) can be solved successively, starting with $i=u$ and
progressing to lower values of $i$.

\subsection{The Bullwhip Effect}\label{Stop}

\begin{figure}[htbp]
\begin{center}
\end{center}
\vspace*{8pt} \caption[]{Frequency response for different
$\beta_i$ and $T_i=\tau_i=\epsilon_i=1$. For small $\beta_i$,
corresponding to a small prognosis time horizon $\Delta t$, a resonance effect
with an amplification factor greater than 1 can be observed.
Perturbations with a frequency $\alpha$ close to the
eigenfrequencies $\omega_i=1/\sqrt{T_i\tau_i}$ are amplified and
cause variations in stock levels and deliveries to grow along the
supply chain. This is responsible for the bullwhip effect.}
\label{resonance}
\end{figure}

The oscillation amplitude increases from one supplier to the next upstream one, if
\begin{equation}
\frac{f_{i-1}^0}{f_i^0} = \bigg\{ 1 + \frac{\alpha{}^2[\epsilon_i
(\epsilon_i + 2\beta_i) - 2 T_i/\tau_i ] + \alpha{}^4T_i{}^2 }
{(1/\tau_i)^2 + (\alpha\beta_i)^2} \bigg\}^{-1/2} > 1 \, .
\end{equation}
One can see that this resonance effect can occur for $0 < \alpha{}^2 < 2/(T_i\tau_i)
- \epsilon_i(\epsilon_i + 2\beta_i)/T_i^2$.
Therefore, variations in the consumption rate are magnified under the instability condition
\begin{equation}
T_i  > \epsilon_i \tau_i \left( \beta_i + \epsilon_i/2 \right) \, .
\label{eq15}
\end{equation}
\par
Supply chains show the bullwhip effect (which corresponds to the
phenomenon of convective, i.e. upstream amplification),
if the adaptation time $T_i$ is too large, if there is no adaptation to some equilibrium
production speed (corresponding to $\epsilon_i = 0$), or if the production management
reacts too strong to deviations of the actual stock level $N_i$ from the desired one $N_i^0$
(corresponding to a small value of $\tau_i$), see
Fig.~\ref{nettypes}. The latter is very surprising, as it implies
that the strategy
\begin{equation}
 \frac{dQ_i}{dt} = \frac{1}{T_i\tau_i} [ N_i^0 - N_i(t) ] \, ,
\label{eq4}
\end{equation}
which tries to maintain a constant work in progress $N_i(t)=N_i^0$, would ultimately lead to an
undesireable bullwhip effect, given that production units are adjusted individually, i.e. in a
decentralized way. In contrast, the management strategy
\begin{equation}
 \frac{dQ_i}{dt} = \frac{1}{T_i} \bigg\{ - \beta_i \frac{dN_i}{dt}
+ \epsilon_i [Q_i^0 - Q_i(t)] \bigg\}
\end{equation}
would avoid this problem, but it would not maintain a constant work in progress.
\par
\begin{figure}[htbp]
\begin{center}
\end{center}
\vspace*{8pt} \caption[]{(a) Plot of the maximum amplitude of
oscillation in the inventories as a function of the adaptation
time $T$. (b) Plot of the maximum amplitude of the inventories as
a function of the time horizon $\Delta t$ for an adaptation time
of $T=2$. (After Ref.\protect\cite{stabilization})}
\label{nettypes}
\end{figure}
The  control strategy (\ref{eq4}) with a sufficiently large value
of $\tau_i$ would fulfill both requirements. Having the
approximate relation (\ref{mind}) in mind, a forecast with a
sufficiently long prognosis time horizon $\Delta t$
(implying large values of $\beta_i$) is favourable
for production stability. Delays in the determination of the
inventories $N_i$, corresponding to negative values of $\Delta t$ and $\beta_i$,
are destabilizing. Note that the values of $\Delta t$ which are sufficient to
stabilize the system are often much smaller than the adaptation
time $T_i$.\cite{stabilization}

\section{Network Effects (with P\'{e}ter \v{S}eba)}

The question is now, whether and how the bullwhip effect can be generalized to
supply networks and how the dynamic behavior depends on the respective
network structure. Linearization of the adaptation equation (\ref{stra}) leads to
\begin{equation}
 \frac{dq_j}{dt} = - \sum_i V_{ji} n_i(t) - \sum_i W_{ji} \frac{dn_i}{dt} - \sum_l X_{jl} q_l(t) \, .
\label{netcont}
\end{equation}
However, in order to avoid a vast number of parameters $V_{ji}$,
$W_{ji}$, $X_{jl}$ and to gain analytical results, we will assume
product- and sector-independent parameters
$V_{jk} = V \delta_{jk}$, $W_{jk} = W \delta_{jk}$, and $X_{jl} =
\delta_{jl}$ in the following. This case corresponds to the situation that each
production unit $j$ is characterized by one dominating product
$i=j$, which also dominates production control. For this reason,
we additionally set $d_{ij} = \delta_{ij}$. Generalizations are
discussed later (see Sec.~\ref{multigoal}).
\par
Using vector notation, we can write the resulting system of differential equations as
\begin{equation}
 \frac{d\vec{n}}{dt} = \mathbf{S} \vec{q}(t) - \vec{y}(t)
\label{former}
\end{equation}
and
\begin{equation}
 \frac{d\vec{q}}{dt} = - V \vec{n}(t) - W \frac{d\vec{n}}{dt} - \vec{q}(t) \, .
\label{latter}
\end{equation}
Inserting Eq.~(\ref{latter}) into Eq.~(\ref{former}) gives
\begin{equation}
 \frac{d}{dt} \left(
\begin{array}{c}
\vec{n} \\
\vec{q}
\end{array} \right) = \mathbf{M} \left(
\begin{array}{c}
\vec{n} \\
\vec{q}
\end{array} \right) +
\left(
\begin{array}{c}
-\vec{y} \\
W\vec{y}
\end{array} \right)
\label{dgl1}
\end{equation}
with
\begin{equation}
 \mathbf{M} =
\left(
\begin{array}{ccc}
\mathbf{0} & , & \mathbf{S} \\
-V\mathbf{E} & , & -\mathbf{E} - W\mathbf{S}
\end{array} \right)
\label{M}
\end{equation}
and the supply matrix $\mathbf{S} = \mathbf{D} - \mathbf{C} = (
d_{ij} - c_{ij})$. For comparison of Eq.~(\ref{lin1}) the one
above, one has to scale the time by introducing the unit time
$T/\epsilon$ and to set $V = 1/(\tau \epsilon)$,
$W=\beta/\epsilon$.

\subsection{General Methods of Solution}\label{Sol}

It is possible to rewrite the system of $2u$ first order differential equations (\ref{dgl1}) in the form
of a system of $u$ differential equations of second order:
\begin{equation}
 \frac{d^2\vec{q}}{dt^2} + (\mathbf{E} + \mbox{$W$}\mathbf{S}) \frac{d\vec{q}}{dt} 
+ \mbox{$V$}\mathbf{S} \vec{q}(t) = \vec{g}(t) \, , 
\label{dgl}
\end{equation}
where
\begin{equation}
\vec{g}(t) = V \vec{y}(t) + W \frac{d\vec{y}}{dt} \, . 
\end{equation}
Introducing the Fourier transforms
\begin{equation}
 \vec{F}(\alpha) = \frac{1}{\sqrt{2\pi}} \int\limits_{-\infty}^\infty \! dt \; \mbox{e}^{-{\rm i}\alpha t} \vec{q}(t)
\end{equation}
and
\begin{equation}
 \vec{G}(\alpha) = \frac{1}{\sqrt{2\pi}} \int\limits_{-\infty}^\infty  \! dt \; \mbox{e}^{-{\rm i}\alpha t}  \vec{g}(t) 
\end{equation}
reduces the problem to solving 
\begin{equation}
 \left[ - \alpha^2 + {\rm i} \alpha (\mathbf{E} + W\mathbf{S}) + V\mathbf{S} \right]
 \vec{F}(\alpha) = \vec{G}(\alpha)
\end{equation}
with given $\vec{G}(\alpha)$. The variable $\alpha$ represents the perturbation frequencies, 
and the general solution of Eq.~(\ref{dgl}) is given by
\begin{equation}
 \vec{q}(t) = \frac{1}{\sqrt{2\pi}} \int\limits_{-\infty}^\infty \! d\alpha \; 
 \mbox{e}^{{\rm i}\alpha t} \vec{F}(\alpha) \, . 
\end{equation}
\par
Note that there exists a matrix
$\mathbf{T}$ which allows one to transform the matrix $\mathbf{E}- \mathbf{S}$
via $\mathbf{T}^{-1}(\mathbf{E} - \mathbf{S})\mathbf{T} = \mathbf{J}$ into either
a diagonal or a Jordan normal form $\mathbf{J}$.
Defining $\vec{x}(\tau) = \mathbf{T}^{-1}\vec{q}(\tau)$
and $\vec{h}(t) = \mathbf{T}^{-1}\vec{g}(t)$, we obtain the coupled set of second-order
differential equations
\begin{equation}
 \frac{d^2x_i}{dt^2} + 2\gamma_i \frac{dx_i}{dt} + \omega_i{}^2 x_i(t)
=  b_i \left[Wx_{i+1}(t) + V \frac{dx_{i+1}}{dt} \right] + h_i(t)  \, ,
\label{chain}
\end{equation}
where
\begin{equation}
 \gamma_i = [1+W(1-J_{ii})]/2 \, , \quad
 \omega_i = [V(1-J_{ii})]^{1/2} \quad \mbox{and} \quad
 b_i = J_{i,i+1} \, .
\end{equation}
This can be interpreted as a set of equations for linearly
coupled damped oscillators with damping constants $\gamma_i$,
eigenfrequencies $\omega_i$, and external forcing $h_i(t)$.
The other forcing terms on the right-hand side are due to interactions of suppliers.
They appear only if $\mathbf{J}$ is not of diagonal, but of Jordan normal form with some
$J_{i,i+1}\ne 0$. Because of $b_i = J_{i,i+1}$, Eqs. (\ref{chain}) can again be analytically
solved in a recursive way, starting with the highest index $i=u$.
Note that, in the case $\mathbf{D}=\mathbf{E}$ (i.e. $d_{ij}=\delta_{ij}$), $J_{ii}$ are the
eigenvalues of the input matrix $\mathbf{C}$ and $0 \le |J_{ii}| \le 1$.
Equation~(\ref{chain}) has a special periodic solution of the form
\begin{eqnarray}
 x_{i}(t) &=& x_{i}^0 \, \mbox{e}^{{\rm i}(\alpha t - \chi_{i})} \, , \\
 h_i(t) &=& h_i^0 \, \mbox{e}^{{\rm i}\alpha t} \, ,
\label{solution}
\end{eqnarray}
where ${\rm i} = \sqrt{-1}$ denotes the imaginary unit.
Inserting this into (\ref{chain}) and dividing
by $\mbox{e}^{{\rm i}\alpha t}$ immediately gives
\begin{equation}
  (-\alpha^2 + 2{\rm i}\alpha \gamma_i + \omega_i{}^2) x_i^0 \mbox{e}^{-{\rm i}\chi_i}
 = b_i (V + {\rm i}\alpha W) x_{i+1}^0 \mbox{e}^{-{\rm i}\chi_{i+1}} + h_i^0 \, .
\end{equation}
With $\mbox{e}^{\pm{\rm i}\phi} = \cos (\phi) \pm {\rm i} \sin(\phi)$ this implies
\begin{equation}
 x_i^0 \mbox{e}^{-{\rm i}\chi_i}
=  \frac{b_i (V + {\rm i}\alpha W) x_{i+1}^0 \mbox{e}^{-{\rm i}\chi_{i+1}} + h_i^0}
 { -\alpha^2 + 2{\rm i}\alpha \gamma_i + \omega_i{}^2}
= \frac{\sqrt{\mbox{Re}^2+\mbox{Im}^2} \, \mbox{e}^{{\rm i}\rho_i}}
{\sqrt{(\omega_i{}^2 - \alpha^2)^2 + (2\alpha\gamma_i)^2} \, \mbox{e}^{{\rm i}\varphi_i}} \, ,
\end{equation}
where
\begin{eqnarray}
\mbox{Re} &=& b_ix_{i+1}^0 [ V\cos(\chi_{i+1})+ \alpha W\sin(\chi_{i+1})] + h_i^0 \, , \nonumber \\
\mbox{Im} &=& b_ix_{i+1}^0 [ \alpha W \cos(\chi_{i+1})- V \sin(\chi_{i+1})] \, ,
\end{eqnarray}
and
\begin{equation}
 x_i^0 = \sqrt{ \frac{[V^2+(\alpha W)^2](b_ix_{i+1}^0)^2 + h_i^0 H_i + (h_i^0)^2 }
{(\omega_i{}^2 - \alpha^2)^2 + (2\alpha\gamma_i)^2} }
\label{help}
\end{equation}
with
\begin{equation}
H_i = 2 b_i x_{i+1}^0 [V \cos(\chi_{i+1}) + \alpha W \sin(\chi_{i+1})] \, .
\end{equation}
Finally, we have
\begin{equation}
 \chi_i = \varphi_i - \rho_i \quad  \mbox{with} \quad \tan \varphi_i
= 2\alpha \gamma_i/ (\omega_i{}^2 - \alpha^2)
\end{equation}
and
\begin{equation}
 \tan \rho_i = \frac{b_ix_{i+1}^0 [ \alpha W \cos(\chi_{i+1})- V \sin(\chi_{i+1})]}
{ b_ix_{i+1}^0 [V \cos(\chi_{i+1})+ \alpha W \sin(\chi_{i+1})] + h_i^0} \, .
\end{equation}
For $h_i^0 = 0$, we obtain
\begin{equation}
\tan (\varphi_i - \chi_i) = \tan(\delta - \chi_{i+1})
\end{equation}
with
\begin{equation}
 \tan \delta = \alpha W/V \, ,
\end{equation}
i.e. the phase shift between $i$ and $i+1$ is just
\begin{equation}
 \chi_i - \chi_{i+1} = \varphi_i - \delta \, .
\end{equation}
\par
According to Eq.~(\ref{chain}), {\em the dynamics of our supply network
model can be reduced to the dynamics of a sequential supply chain.
However, the eigenvalues are now potentially complex and the new
entities $i$ have the meaning of ``quasi-suppliers''} (analogously
to ``quasi-species''\cite{Eigen,Padgett}) defined by the linear
combination $\vec{x}(\tau)= \mathbf{T}^{-1} \vec{q}(\tau)$. This
transformation makes it possible to define the bullwhip effect
for arbitrary supply networks: It occurs if the amplitude
$x_i^0$ is greater than the oscillation amplitude $x_{i+1}^0$
of the next downstream supplier $i+1$ and greater than the
amplitude $h_i^0$ of the external forcing, i.e. if
$x_i^0/\max(x_{i+1}^0,h_i^0) > 1$.
\par
Note that in contrast to sequential supply chains, the oscillation amplitude of $\vec{x}(t)$ may
be amplified in the course of time, depending on the network structure.
This case of {\em absolute instability} can occur
if at least one of the eigenvalues $\lambda_{i,\pm}$ of the homogeneous equation
(\ref{chain}) resulting for $h_i = b_i = 0$ has a positive real part, which may be true when some
complex eigenvalues $J_{ii}$ exist. The (up to) $2u$ eigenvalues
\begin{eqnarray}
 \lambda_{i,\pm}
&=& -\gamma_i \pm \sqrt{\gamma_i{}^2 - \omega_i{}^2} \nonumber \\
&=& -[1+W(1-J_{ii})]/2 \pm \sqrt{ [1+W(1-J_{ii})]^2/4 - V(1-J_{ii})} \qquad
\label{omega2}
\end{eqnarray}
depend on the (quasi-)supplier $i$ and determine the temporal evolution of the amplitude of
deviations from the stationary solution.

\subsection{Examples of Supply Networks}\label{examples}
It is useful to distinguish the following cases:

{\noindent\bf Symmetric supply networks:} If the supply matrix
$\mathbf{S}$ is symmetric, as for most mechanical or electrical
oscillator networks, all eigenvalues $J_{ii}$ are real.
Consequently, if $\omega_i < \gamma_i$ (i.e. if $V$ is small
enough), the eigenvalues $\lambda_{i,\pm}$ of $\mathbf{M}$ are
real and negative, corresponding to an overdamped behavior.
However, if $W$ is too small, the system behavior may be
characterized by damped oscillations.

{\noindent\bf Irregular supply networks:} Most natural and
man-made supply networks have directed links, and $\mathbf{S}$ is
not symmetric. Therefore, some of the eigenvalues $J_{ii}$ will
normally be complex, and an overdamped behavior is untypical. The
characteristic behavior is rather of oscillatory nature (although
asymmetry does not always imply complex eigenvalues\cite{mitWitt},
see Fig.~\ref{behaviors}). For small values of $W$, it can even
happen that the real part of an eigenvalue $\lambda_{i,\pm}$
becomes positive. This implies an amplification of
oscillations in time (until the oscillation amplitude is limited
by non-linear terms). Surprisingly, this also applies to most upper
triangular matrices, i.e. when no loops in the material flows
exist.

{\noindent\bf Regular supply networks:} Another relevant case are
regular supply networks. These are mostly characterized by degenerate zero
eigenvalues $J_{ii}=0$ and Jordan normal forms $\mathbf{J}$, i.e. the existence of
non-vanishing upper-diagonal elements $J_{i,i+1}$. Not only
sequential supply chains, but also fully connected graphs, regular
supply ladders, and regular distribution systems belong to this
case\cite{seba} (see Fig.~\ref{fig:regular}a, c, d). This is characterized by the two
$u$-fold degenerate eigenvalues
\begin{equation}
 \lambda_{\pm} = -(1+W)/2 \pm \sqrt{(1+W)^2/4 - V} \, ,
\end{equation}
independently of the suppliers $i$. For small enough values $V<(1+W)^2/4$, the
corresponding supply systems show overdamped behavior, otherwise damped oscillations.
\par\begin{figure}[h] \begin{center}
\end{center}
\caption[]{Different examples of supply networks: (a) Sequential
supply chain, (b) closed supply chain or supply circle, (c)
regular supply ladder, (d) regular hierarchical distribution
network.} \label{fig:regular}
\end{figure}
For the purpose of illustration, the following equations display some regular input matrices $\mathbf{C}$
and their corresponding Jordan matrices $\mathbf{J}$: For a fully connected network we have
\begin{equation}
\mathbf{C} = \left(
\begin{array}{ccccccc}
\frac{1}{u} & \frac{1}{u} & \frac{1}{u} & \frac{1}{u} & \frac{1}{u} & \dots & \frac{1}{u} \\[1.2mm]
\frac{1}{u} & \frac{1}{u} & \frac{1}{u} & \frac{1}{u} & \frac{1}{u} & \dots & \frac{1}{u} \\[1.2mm]
\frac{1}{u} & \frac{1}{u} & \frac{1}{u} & \frac{1}{u} & \frac{1}{u} & \dots & \frac{1}{u} \\[1.2mm]
\frac{1}{u} & \frac{1}{u} & \frac{1}{u} & \frac{1}{u} & \frac{1}{u} & \dots & \frac{1}{u} \\[1.2mm]
\frac{1}{u} & \frac{1}{u} & \frac{1}{u} & \frac{1}{u} & \frac{1}{u} & \dots & \frac{1}{u} \\
\vdots & \vdots & \vdots & \vdots & \vdots & \ddots & \vdots \\
\frac{1}{u} & \frac{1}{u} & \frac{1}{u} & \frac{1}{u} & \frac{1}{u} & \dots & \frac{1}{u} \\[1.2mm]
\end{array}
\right) \quad \mbox{and}  \quad
\mathbf{J} = \left(
\begin{array}{ccccccc}
0 & 0 & 0 & 0 & 0 & \dots & 0 \\
0 & 1 & 0 & 0 & 0 & \dots & 0 \\
0 & 0 & 0 & 0 & 0 & \dots & 0 \\
0 & 0 & 0 & 0 & 0 & \dots & 0 \\
0 & 0 & 0 & 0 & 0 & \dots & 0 \\[-1mm]
\vdots & \vdots & \vdots & \vdots & \vdots & \ddots & \vdots \\
0 & 0 & 0 & 0 & 0 & \dots & 0 \\
\end{array}
\right) \, .
\end{equation}
The Jordan normal matrix $\mathbf{J}$ of a sequential supply chain
corresponds to the input matrix $\mathbf{C}$ itself, i.e.
$\mathbf{J} = \mathbf{C}$. This, however, is quite exceptional.
For the supply ladder displayed in Fig.~\ref{fig:regular}c we have
\begin{equation}
\mathbf{C} = \left(
\begin{array}{cccccccccc}
0 & 0 & \frac{1}{2} & \frac{1}{2} & 0 & 0 & 0 & 0 & 0 & 0 \\[1mm]
0 & 0 & \frac{1}{2} & \frac{1}{2} & 0 & 0 & 0 & 0 & 0 & 0 \\[1mm]
0 & 0 & 0 & 0 & \frac{1}{2} & \frac{1}{2} & 0 & 0 & 0 & 0 \\[1mm]
0 & 0 & 0 & 0 & \frac{1}{2} & \frac{1}{2} & 0 & 0 & 0 & 0 \\[1mm]
0 & 0 & 0 & 0 & 0 & 0 & \frac{1}{2} & \frac{1}{2} & 0 & 0 \\[1mm]
0 & 0 & 0 & 0 & 0 & 0 & \frac{1}{2} & \frac{1}{2} & 0 & 0 \\[1mm]
0 & 0 & 0 & 0 & 0 & 0 & 0 & 0 & \frac{1}{2} & \frac{1}{2} \\[1mm]
0 & 0 & 0 & 0 & 0 & 0 & 0 & 0 & \frac{1}{2} & \frac{1}{2} \\[1mm]
0 & 0 & 0 & 0 & 0 & 0 & 0 & 0 & 0 & 0 \\[1mm]
0 & 0 & 0 & 0 & 0 & 0 & 0 & 0 & 0 & 0
\end{array}
\right) \quad \mbox{and} \quad
\mathbf{J} = \left(
\begin{array}{cccccccccc}
0 & 1 & 0 & 0 & 0 & 0 & 0 & 0 & 0 & 0 \\
0 & 0 & 1 & 0 & 0 & 0 & 0 & 0 & 0 & 0 \\
0 & 0 & 0 & 1 & 0 & 0 & 0 & 0 & 0 & 0 \\
0 & 0 & 0 & 0 & 1 & 0 & 0 & 0 & 0 & 0 \\
0 & 0 & 0 & 0 & 0 & 0 & 0 & 0 & 0 & 0 \\
0 & 0 & 0 & 0 & 0 & 0 & 0 & 0 & 0 & 0 \\
0 & 0 & 0 & 0 & 0 & 0 & 0 & 0 & 0 & 0 \\
0 & 0 & 0 & 0 & 0 & 0 & 0 & 0 & 0 & 0 \\
0 & 0 & 0 & 0 & 0 & 0 & 0 & 0 & 0 & 0 \\
0 & 0 & 0 & 0 & 0 & 0 & 0 & 0 & 0 & 0 \\
\end{array}
\right) \, ,
\end{equation}
where the number of ones corresponds to the number of levels of the supply ladder.
For the hierarchical distribution network shown in Fig.~\ref{fig:regular}d, but with
3 levels only, we have
\begin{equation}
\mathbf{C} = \left(
\begin{array}{ccccccc}
0 & \frac{1}{2} & \frac{1}{2} & 0 & 0 & 0 & 0 \\[1mm]
0 & 0 & 0 & \frac{1}{2} & \frac{1}{2} & 0 & 0 \\[1mm]
0 & 0 & 0 & 0 & 0 & \frac{1}{2} & \frac{1}{2} \\[1mm]
0 & 0 & 0 & 0 & 0 & 0 & 0 \\
0 & 0 & 0 & 0 & 0 & 0 & 0 \\
0 & 0 & 0 & 0 & 0 & 0 & 0 \\
0 & 0 & 0 & 0 & 0 & 0 & 0 \\
\end{array}
\right) \quad \mbox{and} \quad \mathbf{J} = \left(
\begin{array}{ccccccc}
0 & 1 & 0 & 0 & 0 & 0 & 0 \\
0 & 0 & 1 & 0 & 0 & 0 & 0 \\
0 & 0 & 0 & 0 & 0 & 0 & 0 \\
0 & 0 & 0 & 0 & 1 & 0 & 0 \\
0 & 0 & 0 & 0 & 0 & 0 & 0 \\
0 & 0 & 0 & 0 & 0 & 0 & 1 \\
0 & 0 & 0 & 0 & 0 & 0 & 0
\end{array}
\right) \, .
\label{JHier}
\end{equation}
Note that the Jordan normal forms of different input matrices $\mathbf{C}$ may be identical,
but the transformation matrix $\mathbf{T}$ and the driving term $\vec{h}(t)$
would then be different.

{\noindent\bf Randomized regular supply networks} belong to the
class of irregular supply networks, but they can be viewed as
slightly perturbed regular supply networks. For this reason, there
exist approximate analytical results for their eigenvalues. Even
very small perburbations of the regular matrices $\mathbf{S}$
discussed in the previous paragraph can change the eigenvalue
spectrum qualitatively. Instead of the two multiply degenerate
eigenvalues $\lambda_{\pm}$ of Eq.~(\ref{omega2}), we find a
scattering of eigenvalues around these values. The question is:
why?
\par
In order to assess the behavior of randomized regular supply
networks, we apply Ger\v{s}gorin's theorem on the location of
eigenvalues.\cite{Matrix} According to this, the $n$ complex
eigenvalues $\lambda_k\in \mathbb{C}$ of some $n\times n$-matrix
$\mathbf{N}$ are located in the union of $n$ disks:
\begin{equation}
 \lambda_k \in \bigcup_i \left\{ z \in \mathbb{C}: |z - N_{ii}| \le \sum_{j(\ne i)} |N_{ij}| \right\}
\end{equation}
Furthermore, if a union of $l$ of these $n$ discs form a connected region that is disjoint from
all the remaining $n-l$ discs, then tere are precisely $l$ eigenvalues of $\mathbf{N}$ in this region.
\par
Let us now apply this theorem to the perturbed matrix
\begin{equation}
\mathbf{C}_\eta = \mathbf{C} + \eta \mathbf{P} \, .
\end{equation}
For small enough values of $\eta$, {\em the corresponding eigenvalues $J_{ii}$ should be
located within discs of radius
\begin{equation}
 R_i(\eta) = \eta \sum_{j} |P_{ij}^{(\eta)}|
\end{equation}
around the (possibly degenerated) eigenvalues $J_{ii}$ of the
original matrix $\mathbf{C}$} (see Fig.~\ref{perturb}). This radius grows monotonously, but not 
necessarily linearly in the parameter $\eta$ with $0 < \eta \le 1$, which allows to control the size
of the perturbation.  Moreover, $\mathbf{P}^{(\eta)} =
\mathbf{R}_{\eta}^{-1} \mathbf{P} \mathbf{R}_{\eta}$, where
$\mathbf{R}_{\eta}$ is the orthogonal matrix which transforms
$\mathbf{C}_\eta$ to a diagonal matrix $\mathbf{D}^{(\eta)}$, i.e.
$\mathbf{R}_{\eta}^{-1} \mathbf{C}_\eta  \mathbf{R}_{\eta} =
\mathbf{D}^{(\eta)}$. (This assumes a perturbed matrix
$\mathbf{C}_\eta$ with no degenerate eigenvalues.) Similar discs
as for the eigenvalues of $\mathbf{C}_\eta$ can be determined for
the associated eigenvalues $\lambda_{i,\pm}^{(\eta)}$ of the
perturbed $(2u\times 2u)$-matrix $\mathbf{M}_\eta$ belonging to
the perturbed $u\times u$-matrix $\mathbf{C}_\eta$, see
Eq.~(\ref{M}) and Fig.~\ref{perturb}.
\par
\begin{figure}[htbp]
\begin{center}
\end{center}
\vspace*{8pt} \caption[]{(a) Example of a supply network
(regular distribution network) with two degenerated real
eigenvalues (see squares in the right subfigure). (b) The
eigenvalues $\lambda_{i,\pm}$ of the randomly perturbed supply
network (crosses) are mostly complex and located within
Ger\v{s}gorin's discs (large circles). (After
Ref.\protect\cite{seba})} \label{perturb}
\end{figure}
Let us now discuss the example of a structural perturbation of a
sequential supply chain (with $\eta = 0$) towards a supply circle
(with $\eta = 1$), see Fig.~\ref{fig:regular}b. For this, we set
\begin{equation}
\mathbf{C} = \left(
\begin{array}{ccccc}
0 & 1 & 0 & \ldots & 0 \\
0 & 0 & 1 & \ldots & 0 \\[-1mm]
\vdots  & &  & \ddots & \vdots \\
0 & 0 & 0 & \ldots & 1 \\
\eta\; & 0\; & 0\; & \ldots & 0
\end{array}
\right) \, .
\end{equation}
While the normal form for $\eta = 0$ is given by a Jordan matrix
$\mathbf{J}$ which agrees with $\mathbf{C}$, for any $\eta > 0$ we find the
diagonal matrix
\begin{equation}
\mathbf{J} = \left(
\begin{array}{cccc}
J_{11} & 0 & \ldots & 0 \\
0 & J_{22} & \ldots & 0 \\
\vdots  &  & \ddots & \vdots \\
0 & 0 & 0 & J_{nn}  
\end{array}
\right) \, ,
\label{circle}
\end{equation}
where the diagonal elements $J_{ii}$ are complex and equally distributed on a circle
of radius $\sqrt[n]{\eta}$ around the origin of the complex plane.
Therefore, even an arbitrarily small perturbation can change the
eigenvalues qualitatively and remove the degeneration of the
eigenvalues.

\section{Network-Induced Business Cycles (with Ulrich Witt and Thomas
Brenner)}\label{businesscycles}

In order to investigate the macroeconomic dynamics of national economies, it is useful to
identify the production units with economic sectors (see Fig.~\ref{sectors}). It is also
necessary to extend the previous supply network model by price-related equations.
For example, increased prices $P_i(t)$ of products of sector $i$
have a negative impact on the consumption rate $Y_i(t)$
and vice versa. We will describe this by a standard demand function $L_i$ with
a negative derivative $L'_i(P_i)= dL_i(P_i)/dP_i$:
\begin{equation}
 Y_i(t) = [Y_i^0 + \xi_i(t)] L_i(P_i(t)) \, .
\label{Y}
\end{equation}
This formula takes into account
random fluctuations $\xi_i(t)$ over time around a certain average consumption rate $Y_i^0$ and assumes
that the average value of $L_i(P_i(t))$ is normalized to one. The fluctuation term $\xi_i(t)$ is introduced
here in order to indicate that the variation of the consumption rate is a potentially relevant source of
fluctuations.
\par\begin{figure}[htbp]
\begin{center}
\end{center}
\vspace*{8pt} \caption[]{Main service and commodity flows among
different economic sectors according to averaged input-output data
of France, Germany, Japan, UK, and USA. For clarity of the
network structure, we have omitted the sector `wholesale and
retail trade', which is basically connected with all other
sectors.} \label{sectors}
\end{figure}
Inserting (\ref{Y}) into (\ref{conserv}) results in
\begin{equation}
\frac{dN_i(t)}{dt} = Q_i(t) - \sum_{j=1}^m c_{ij} Q_j(t) -
\underbrace{[Y_i^0 + \xi_i(t)] L_i(P_i(t))}_{=Y_i(t)} \,
\label{N}.
\end{equation}
Herein, we have assumed $d_{ij} = \delta_{ij}$ as in Leontief's
classical input-output model.\cite{Leontief} Moreover, in our
simulations we have applied the common linear demand function
\begin{equation}
    L_i(P_i) = \max(0,L_i^0 - L_i^1 P_i) \, ,
\label{F}
\end{equation}
where $L_i^0$ and $L_i^1$ are non-negative parameters.
\par
Due to the price dependence of consumption,
economic systems have the important equilibration mechanism of price adjustment.
These can compensate for undesired inventory levels and inventory changes. A mathematical
equation reflecting this is
\begin{equation}
  \frac{1}{P_i(t)} \frac{dP_i}{dt} = \nu_i \left( \frac{N_i^0}{N_i(t)} - 1 \right) -
 \frac{\mu_i}{N_i(t)} \frac{dN_i}{dt} \, .
\label{P}
\end{equation}
The use of relative changes guarantees the required non-negativity of prices $P_i(t) \ge 0$.
$\nu_i$ is an adaptation rate describing the sensitivity to relative deviations of the actual
inventory $N_i(t)$ from the desired one $N_i^0$, and $\mu_i$ is a dimensionless parameter
reflecting the responsiveness to relative deviations $(dN_i/dt)/N_i(t)$ from the stationary
equilibrium state.
\par
If the same criteria are applied to adjustments of the production rates $Q_i(t)$, we
have the equation
\begin{equation}
  \frac{1}{Q_i(t)} \frac{dQ_i}{dt} = \hat{\nu}_i
\left( \frac{N_i^0}{N_i(t)} - 1 \right) - \frac{\hat{\mu}_i}{N_i(t)}
\frac{dN_i}{dt} \, .
\label{Q}
\end{equation}
$\hat{\alpha}_i = \hat{\nu}_i/\nu_i$ is the ratio between the adjustment
rate of the output flow and the adjustment rate of the price in sector $i$. For simplicity,
the same ratio $\hat{\alpha}_i = \hat{\mu}_i/\mu_i$ will be assumed for the responsiveness.

\subsection{Treating Producers Analogous to Consumers (with Dieter Armbruster)}

According to Eqs.~(\ref{Y}) and (\ref{F}), the change of the consumption rate in time is basically given by
\begin{equation}
 \frac{dY_i}{dt} = [Y_i^0 + \xi_i(t)] \frac{dL_i}{dP_i} \frac{dP_i}{dt} = - [Y_i^0 + \xi_i(t)] L_i^1 \frac{dP_i}{dt} \, ,
\label{dY}
\end{equation}
i.e. it is basically proportional to the price change $dP_i/dt$, but with a {\em negative} and
potentially fluctuating prefactor. However,
\begin{equation}
 \frac{1}{Q_i(t)} \frac{dQ_i}{dt} \propto \frac{1}{P_i(t)} \frac{dP_i}{dt} \, ,
\end{equation}
i.e. according to Eqs.~(\ref{P}) and (\ref{Q}), the change $dQ_i/dt$ of the production rate close to the
equilibrium state with $Q_i(t) \approx Q_i^0$ and $P_i(t) \approx P_i^0$ is basically proportional
to the change $dP_i/dt$ in the price with a {\em positive} prefactor. Is this an inconsistency of the model?
Shouldn't we better treat producers analogous to consumers?
\par
In order to do so, we have to introduce the delivery flows $D_{ij}$ of products $i$ to producers $j$.
As in Eq.~(\ref{dY}), we will assume that their change $dD_{ij}/dt$ in time is proportional to the
change $dP_i/dt$ in the price, with a negative (and potentially fluctuating) prefactor. We have now
to introduce the stock level $O_i(t)$ in the output buffer of producer $i$ and the stock levels
$I_{ij}(t)$ of product $i$ in the input buffers of producer $j$. For the output buffer, we find the
balance equation
\begin{equation}
 \frac{dO_i}{dt} = Q_i(t) - \sum_j D_{ij}(t) - Y_i(t)
\end{equation}
analogous to Eq.~(\ref{N}), as the buffer is filled with the production rate of producer $i$, but is emptied by
the consumption rate and delivery flows. For
the input buffers we have the balance equations
\begin{equation}
 \frac{dI_{ij}}{dt} = D_{ij}(t) - c_{ij} Q_j(t) \, ,
\end{equation}
as these are filled by the delivery flows, but emptied with a rate proportional to the production rate
of producer $j$, where $c_{ij}$ are again the input coefficients specifying the relative quantities of required
inputs $i$ for production. Generalizations of these equations are discussed elsewhere.\cite{NJP}
\par
Let us now investigate the differential equation for the change $dN_i/dt$ of the overall stock level of
product $i$ in all input and output buffers. We easily obtain
\begin{equation}
 \frac{dN_i}{dt} = \frac{dO_i}{dt} + \sum_j \frac{dI_{ij}}{dt} = Q_i(t) - \sum_j c_{ij} Q_j(t) - Y_i(t) \, ,
\end{equation}
as before. That is, the equations for the delivery flows drop out, and we stay with the previous
set of equations for $N_i$, $Q_i$, $P_i$, and $Y_i$. As a consequence, we will focus on
Eqs.~(\ref{Y}) to (\ref{Q}) in the following.

\section{Reproduction of Some Empirically Observed Features of Business Cycles} \label{Bus}

Our simulations of the above dis-aggregate (i.e. sector-wise)
model of macroeconomic dynamics typically shows asynchronous
oscillations, which seems to be characteristic for economic
systems. Due to phase shifts between sectoral oscillations, the
aggregate behavior displays slow variations of small amplitude
(see Fig.~\ref{fig:business}). If
the function $L_i(P_i)$ and the parameters $\nu_i/\mu_i{}^2$ are
suitably specified, the non-linearities in Eqs.~(\ref{N}) to
(\ref{Y}) will additionally limit the oscillation amplitudes.
\par\begin{figure}[htbp]
\begin{center}
\end{center}
\vspace*{8pt} \caption[]{Typical simulation result of the
time-dependent gross domestic product $\sum_i Q_i(t) P_i(t)$ in
percent, i.e. relative to the initial value. The input matrix was
chosen as in Figs.~\ref{sectors} and
\ref{behaviors}a--d, but $Y_i^0$ was determined from averaged
input-output data. $Q_i^0$ was obtained from the equilibrium
condition, and the fluctuations $\xi_i(t)$ were specified as a
Gaussian white noise with mean value 0 and variance $\sigma =
10.000$ (about 10\% of the average final consumption).  The
initial prices $P_i(0)$ were selected from the interval [0.9;1.1].
Moreover, in this example we have assumed $L_i(P_i) =
\max[0,1+d(P_i-P_i^0)]$ with $d=f'(P_i^0) = - L_i^1 = -10$ and the
parameters $\nu_i = 0.1$, $\mu_i = 0.0001$, $\alpha_i = 1 =
P_i^0$, and $N_i^0 = Y_i^0$.   Although this implies a growth of
small oscillations (cf. Fig.~1d), the oscillation amplitudes are
rather limited. This is due to the non-linearity of model
equations (\ref{N}) to (\ref{Y}) and due to the phase shifts
between oscillations of different economic sectors $i$. Note that
irregular oscillations with frequencies between 4 to 6 years and
amplitudes of about 2.5\% are qualitatively well compatible with
empirical business cycles. Our material flow model can explain
$w$-shaped, non-periodic oscillations without having to assume
technological shocks or externally induced perturbations. The
long-term growth of national economies was intentionally not
included in the model in order to separate this effect from
network-induced instability effects. (After
Ref.\protect\cite{mitWitt})} \label{fig:business}
\end{figure}
Our business cycle theory differs from the dominating
one\cite{econo1,econo2} in several favourable aspects:
\begin{itemize}
\item[(i)] Our theory explains irregular, i.e. non-periodic
oscillations in a natural way (see Fig.~\ref{fig:business}). For
example, $w$-shaped oscillations result as superposition of the
asynchronous oscillations in the different economic sectors, while
other theories have to explain this observation by assuming
external perturbations  (e.g. due to technological innovations).
\item[(ii)] Although our model may be extended by variables such
as the labor market, interest rates, etc., we consider it as a
potential advantage that we did not have to couple variables in
our model which are qualitatively that different. Our model rather
focusses on the material flows among different sectors.
\item[(iii)] Moreover, we will see that our model can explain
emergent oscillations, which are not triggered by external shocks.
\end{itemize}
\begin{figure}[htbp]
\begin{center}
\end{center}
\vspace*{8pt} \caption[]{The displayed phase diagram shows which
dynamic behavior is expected by our dis-aggregate model of
macroeconomic dynamics depending on the respective parameter
combinations. The individual curves for different countries are a
result of the different structures of their respective input
matrices $\mathbf{C}$. As a consequence, structural policies can
influence the stability and dynamics of economic systems.}
\label{countries}
\end{figure}

\subsection{Dynamic Behaviors and Stability Thresholds}

The possible dynamic behaviors of the resulting dis-aggregate
macroeonomic model can be studied by analytical investigation of
limiting cases and by means of a linear stability analysis around
the equilibrium state (in which we have $N_i(t) = N_i^0$, $Y_i(t)
= Y_i^0$, $Q_i^0 - \sum_j c_{ij} Q_j^0 = Y_i^0$, and $P_i(t) =
P_i^0$). The linearized equations for the deviations $n_i(t) =
N_i(t) - N_i^0$, $p_i(t) = P_i(t) - P_i^0$, and $q_i(t) = Q_i(t) -
Q_i^0$ from the equilibrium state read
\begin{eqnarray}
 \frac{dn_i}{dt} &=& q_i - \sum_j c_{ij} q_j - Y_i^0 f'_i(P_i^0) p_i - \xi_i(t) \, ,
\label{n} \\
 \frac{dp_i}{dt}
 &=& \frac{P_i^0}{N_i^0} \left( - \nu_i n_i  - \mu_i \frac{dn_i}{dt} \right) \, ,
\label{p} \\
 \frac{dq_i}{dt}
&=&  \frac{\hat{\alpha}_i Q_i^0}{N_i^0} \left( - \nu_i n_i  - \mu_i  \frac{dn_i}{dt} \right) \, .
\label{q}
\end{eqnarray}
This system of coupled differential equations describes the response of the inventories,
prices, and production rates to variations $\xi_i(t)$ in the demand.
Denoting the $m$ eigenvalues
of the input matrix $\mathbf{C}= (c_{ij})$ by $J_{ii}$ with $|J_{ii}| < 1$,
the $3m$  eigenvalues of the linearized model equations are $0$ ($m$ times) and
\begin{equation}
 \lambda_{i,\pm} \approx \frac{1}{2} \bigg( - A_i \pm \sqrt{ (A_i)^2 - 4 B_i} \, \bigg) \, ,
\label{EV}
\end{equation}
where
\begin{eqnarray}
A_i &=& \mu_i [C_i + \hat{\alpha}_i D_i(1-J_{ii})] \, , \nonumber \\
B_i &=& \nu_i [C_i + \hat{\alpha}_i D_i(1-J_{ii})] \, , \nonumber \\
C_i &=& P_i^0 Y_i^0 |f'_i(P_i^0)| /N_i^0 \, ,  \nonumber \\
D_i &=& Q_i^0 /N_i^0 \, ,
\end{eqnarray}
see Fig.~\ref{countries}. Formula (\ref{EV}) becomes exact when the matrix $\mathbf{C}$
is diagonal or the parameters $\mu_i C_i$, $\hat{\alpha}_i \mu_i D_i$, $\nu_i C_i$ and $\hat{\alpha}_i \nu_i D_i$
are sector-independent constants, otherwise the eigenvalues must be numerically determined.
\par
\begin{figure}[htbp]
\begin{center}
\end{center}
\vspace*{8pt} \caption[]{Properties of our dynamic model of supply
networks for a characteristic input matrix specified as average
input matrix of macroeconomic commodity flows of several countries
(top) and for a synthetic input matrix generated by random changes
of input matrix entries until the number of complex eigenvalues
was eventually reduced to zero (bottom). Subfigures {(a), (e)}
illustrate the color-coded input matrices $\mathbf{A}$, {(b), (f)}
the corresponding network structures, when only the strongest
links (commodity flows) are shown, {(c), (g)} the eigenvalues
$J_{ii}=\mbox{Re}(J_{ii}) + \mbox{i} \,\mbox{Im}(J_{ii})$ of
the respective input matrix $\mathbf{A}$, and {(d), (h)} the phase
diagrams indicating the stability behavior of the model equations
(\ref{N}) to (\ref{Y}) on a double-logarithmic scale as a function
of the model parameters $\hat{\alpha}_i = \hat{\alpha}$ and $\nu_i/\mu_i{}^2 =
\nu/\mu^2 = V/M^2$. The other model parameters were set to $\nu_i
= C_i = D_i = P_i^0 =N_i^0 = Y_i^0 = 1$. Surprisingly, for
empirical input matrices $\mathbf{A}$, one never finds an
overdamped, exponential relaxation to the stationary equilibrium
state, but network-induced oscillations due to complex eigenvalues
$J_{ii}$. (After Ref.\protect\cite{mitWitt})} \label{behaviors}
\end{figure}
It turns out that the dynamic behavior mainly depends on the
parameters $\hat{\alpha}_i$, $\nu_i/\mu_i{}^2$, and the
eigenvalues $J_{ii}$ of the input matrix $\mathbf{C}$ (see
Fig.~\ref{behaviors}): In the case $\hat{\alpha}_i \rightarrow 0$
of fast price adjustment, the eigenvalues $\lambda_{i,\pm}$ are
given by
\begin{equation}
 2\lambda_{i,\pm} = - \mu_iC_i \pm \sqrt{ (\mu_i C_i) ^2 - 4 \nu_i C_i} \, ,
\end{equation}
i.e. the network structure does not matter at all. We expect an
exponential relaxation to the stationary equilibrium for $0 <
\nu_i/ \mu_i{}^2 < C_i/4$, otherwise damped oscillations.
Therefore, immediate price adjustments or similar mechanisms are
an efficient way to stabilize economic and other supply systems.
However, any delay ($\hat{\alpha}_i > 0$) will cause damped or
growing oscillations, if complex eigenvalues  $J_{ii} =
\mbox{\rm Re}(J_{ii}) + \mbox{\rm i\,Im}(J_{ii})$ exist.
\textit{Note that this is the normal case, as typical supply
networks in natural and man-made systems are characterised by
complex eigenvalues} (see top of Fig.~\ref{behaviors}).
\par
Damped oscillations can be shown to result if all values
\begin{equation}
\nu_i/\mu_i{}^2 = \hat{\alpha}_i \hat{\nu}_i/\hat{\mu}_i{}^2
\end{equation}
lie below the instability lines
\begin{eqnarray}
\nu_i/\mu_i{}^2 &\approx& \Big\{C_i + \hat{\alpha}_i D_i [1-\mbox{Re}(J_{ii})]\Big\} \nonumber \\
&\times& \left( 1 + \frac{\{C_i+\hat{\alpha}_i D_i [1-\mbox{Re}(J_{ii})]\}^2}
{[\hat{\alpha}_i D_i\mbox{Im}(J_{ii})]^2} \right)
\label{new}
\end{eqnarray}
given by  the condition $\mbox{Re}(\lambda_{i,\pm}) \le 0$. For
identical parameters $\nu_i/\mu_i{}^2 = \nu/\mu^2$ and
$\hat{\alpha}_i = \hat{\alpha}$, the minimum of these lines agrees
exactly with the numerically obtained curve in
Fig.~\ref{behaviors}. Values above this line cause small
oscillations to grow over time (see Appendix~A).
\par
In some cases, all eigenvalues $J_{ii}$ of the input matrix $\mathbf{C}$ are real.
This again applies to symmetric matrices $\mathbf{C}$ and matrices equivalent to Jordan
normal forms. Hence, the existence of loops in
supply networks is no sufficient condition for complex eigenvalues $J_{ii}$ (see also Fig.~1f). It is also no
necessary condition. For cases with real eigenvalues only,
Eq.~(\ref{EV}) predicts a stable, overdamped behavior if
all values $\nu_i/\mu_i{}^2  = \hat{\alpha}_i \hat{\nu}_i/\hat{\mu}_i{}^2$ lie below the lines
\begin{equation}
 \nu_i/\mu_i{}^2 \approx  [ C_i + \hat{\alpha}_i D_i (1-J_{ii})]/4
\label{im}
\end{equation}
defined by $\min_i (A_i{}^2 - 4B_i) > 0$ (see Appendix B). For
identical parameters $\nu_i/\mu_i{}^2 = \nu/\mu^2$ and
$\hat{\alpha}_i = \hat{\alpha}$, the minimum of these lines
corresponds exactly to the numerically determined curve in
Fig.~\ref{behaviors}. Above it, one observes damped oscillations
around the equilibrium state, but growing oscillations are not
possible. In supply systems with a slow or non-existent price adjustment
mechanism (i.e. for $\hat{\alpha}_i \gg 1$
or $C_i = 0$), Eq.~(\ref{im}) predicts an overdamped behavior for
real eigenvalues $J_{ii}$ and
\begin{equation}
 \hat{\nu}_i/\hat{\mu}_i{}^2  < D_i(1-J_{ii}) / 4
\end{equation}
for all $i$, while Eq.~(\ref{new}) implies the stability condition
\begin{equation}
 \hat{\nu}_i/\hat{\mu}_i{}^2
 < D_i [1-\mbox{Re}(J_{ii})] \{1+ [1-\mbox{Re}(J_{ii})]^2/\mbox{Im}(J_{ii})^2\}
\end{equation}
for all $i$, given that some eigenvalues $J_{ii}$ are complex. Moreover, for the case of
sector-independent constants $V = \hat{\alpha}_i \nu_i D_i$ and $W = \hat{\alpha}_i \mu_i D_i$,
the eigenvalues $\lambda_{i,\pm}$ can be calculated as
\begin{equation}
 \lambda_{i,\pm} = - W(1-J_{ii})/2 \pm \sqrt{ [W(1-J_{ii})]^2/4 - V(1-J_{ii})} \, .
\end{equation}
Considering that Eq. (\ref{latter}) for the adjustment of production speeds is slightly more
general than specification (\ref{q}), this result is fully consistent with
Eq.~(\ref{omega2}).

\section{Summary} \label{Sum}

In this contribution, we have investigated a model of supply networks. It is based on
conservation equations describing the storage and flow of inventories by a dynamic variant of
Leontief's classical input-output model. A second set of equations reflects the delayed adaptation
of the production rate to some inventory-dependent desired production rate. Increasing
values of the adaptation time $T$ tend to destabilize the system, while increasing values of
the time horizon $\Delta t$ over which inventories are forecasted have a stabilizing effect.
\par
A linear
stability analysis shows that supply networks are expected to show an overdamped or damped oscillatory
behavior, when the supply matrix is symmetrical. Some regular
supply networks such as sequential supply chains or regular
supply hierarchies with identical parameters show
stable system behavior for all values of the adaptation time $T$. Nevertheless,
one can find the bullwhip effect under certain circumstances, i.e. the oscillations in the
production rates are larger than the variations in the consumption rate (demand). The underlying
mechanism is a resonance effect.
\par
Regular supply networks are characterized by multiply degenerate eigenvalues. When the
related supply matrices are slightly perturbed, the degeneration is
broken up. Instead of two degenerate eigenvalues, one finds complex eigenvalues which
approximately lie on a circle around them in the complex plane. The related heterogeneity
in the model parameters can have a stabilizing effect,\cite{NJP} when they manage to reduce
the resonance effect. Moreover, a randomly perturbed regular supply network may
become linearly unstable, if the perturbation is large. If the supply matrix is irregular,
the supply network is generally expected to show
either damped or growing oscillations. In any case, supply networks can be mapped
onto a generalized sequential supply chain, which allows one to define the bullwhip effect
for supply networks.
\par
While previous studies have focussed on the synchronization of
oscillators in different network topologies,\cite{nets2,PRL} we
have found that \textit{many supply networks display damped
oscillations, even when their units---and linear chains of these
units---behave in an overdamped way. Furthermore, networks of
damped oscillators tend to produce growing (and mostly
asynchronous) oscillations.} Based on these findings, it is
promsing to explain business cycles as a result of material flows
among economic sectors, which adjust their production rates in a
decentralized way. Such a model can be also extended by aspects
like labor force, money flows, information flows, etc.

\section{Future Research Directions}

\subsection{Network Engineering}

Due to the sensitivity of supply systems to their network
structure, network theory\cite{nets1,nets2,nets3,nets4,nets5,nets6,nets7,Stewart} can make
useful contributions: On the basis of Eqs.~(\ref{new}) and
(\ref{im}) one can design stable, robust, and adaptive supply
networks (``network engineering''). For this reason, our present
studies explore the characteristic features of certain types of
networks: random networks, hierarchical networks, small world
networks, preferential attachment networks, and others. In this
way, we want to identify structural measures which can increase
the robustness and reliability of supply networks, also their
failure and attack tolerance. These results should, for example,
be relevant for the optimization of disaster
management.\cite{disaster}

\subsection{Cyclic Dynamics in Biological Systems}

Oscillations are probably not always a bad feature of supply
systems. For example, in systems with competing goals (such as
pedestrian counterflows at bottlenecks\cite{pedestrians} or intersecting traffic
streams), oscillatory solutions can be favourable. Oscillations are also
found in ecological systems\cite{Eco} and
some metabolic processes,\cite{Dane,metabo1,metabo2,metabo3} which could
be also treated by a generalized model of supply networks.
Examples are oscillations of nutrient flows in slime molds,\cite{Ueda}
the citrate cycle,\cite{cytrase} the glycolytic oscillations
in yeast,\cite{Dane} and the Calcium cycle.\cite{calcium1,calcium2,calcium3}
Considering the millions of years of evolution, it is highly
unlikely that these oscillations do not serve certain biological
functions. Apart from metabolic flows, however, we should also
mention protein networks\cite{metabo3} and metabolic expression
networks\cite{metabo1,metabo2} as interesting areas of application of
(generalized) supply network models.

\subsection{Heterogeneity in Production Networks}

Another interesting subject of investigation would be
heterogeneity. The relevance of this issue is known from traffic
flows\cite{heterotraffic} and multi-agent systems in economics\cite{heteroeco}.
Preliminary results indicate that, depending on the network
structure, heterogeneous parameters of production units can reduce
the bullwhip effect in supply networks.\cite{NJP} That is,
identical production processes all over the world may destabilize
economic systems (as monocultures do in agriculture).
However, in heterogeneous systems a lot more frequencies are influencing a
single production step, which may also have undesireable effects.
These issues deserve closer investigation.

\subsection{Multi-Goal Control} \label{multigoal}

Let us go back to Eq.~(\ref{netcont}) for the control of supply or production rates.
This equation is to be applied to cases where a production unit produces several
products and the production rate is not only adapted to the inventory of one
dominating product, but also to the inventories of some other products. This leads to the
problem of multi-goal control. It is obvious that the different goals do not always have to be
in harmony with each other: A low inventory of one product may call for an increase in the production
rate, but the warehouse for another product, which is generated at the same time,
may be already full. A good example is the production of chemicals. We conjecture that
multi-goal control of production networks may imply additional sources of dynamical
instabilities and that some of the problems may be solved by price adjustments,
which can influence the consumption rates. Again, a closer investigation of these
questions is needed.

\subsection{Non-Linear Dynamics and Scarcity of Resources}

Finally, we should note that coupled equations for damped
harmonic oscillators approximate the dynamic behavior of supply
networks only close to their stationary state. When the
oscillation amplitudes become too large, non-linearities will
start to dominate the dynamic behavior. For example, they may
select wave modes or influence their phases in favour of a
synchronized behavior,\cite{NJP,stabilization} etc. Deterministically chaotic
behavior seems to be possible as
well.\cite{beer2,Chaos,Cha3,Cha4,Cha5} Numerical results for
limited buffer sizes and transport capacities have, for example, been presented by
Peters {\em et al.}\cite{Peters}
\par
Additional non-linearities come into play, if there is a scarcity of
resources required to complete certain products. Let's assume that the rate
of transfering products $i$ to $j$ is $Z_{ji}$. If $N_i$ is the number of products
that may be delivered, the maximum delivery rate is therefore limited by
$Z_{ji}N_i(t)$. The maximum production rate is given by the minimum of the
delivery rate of all components $i$, divided by the respective number $c_{ij}$ of
parts required to complete one unit of product $i$. In mathematical terms,
we have to make the replacement
\begin{equation}
 Q_j(t) \longrightarrow
 Q_j(t) \min_i \left(1, \frac{Z_{ji}N_i(t)}{c_{ij}} \right)   
\label{mini}
\end{equation}
in Eq.~(\ref{N}) and corresponding replacements in all derived
formulas.\cite{NJP} Therefore, the generalized relationship for
systems in which resources may run short is
\begin{equation}
 \frac{dN_i}{dt} = \sum_{j=1}^u ( d_{ij} - c_{ij}) Q_j(t)
 \min_k \left(1, \frac{Z_{jk}N_k(t)}{c_{jk}} \right) - Y_i(t) \, .
\end{equation}
These coupled non-linear equations are expected to result
in a complex dynamics, if the transportation rate $Z_{jk}$
or inventory $N_k$ are too small. Note that this non-linearity may be even relevant
close to the stationary state. In such cases, results of linear stability analyses for systems with
scarce resources would be potentially misleading.
\par
In situations where scarce resources may be partially substituted by other available resources, one
may use the replacement
\begin{equation}
 Q_j(t) \longrightarrow
\left[ 1 + \sum_i \left( \frac{Z_{ji} N_i(t)}{c_{ij}} \right)^{z_j} \right]^{1/{z_j}} 
\end{equation}
instead of (\ref{mini}). Such an approach is, for example, reasonable for disaster management.\cite{disaster}
$z_j$ is a parameter allowing to fit the ease of substitution of
resources. The minimum function results in the limit $z_j\rightarrow -\infty$.

\section*{Acknowledgements}
\addcontentsline{toc}{section}{Acknowledgements}

S.L. appreciates a scholarship by the ``Studienstiftung des Deutschen
Volkes''. Moreover, the authors
would like to thank Thomas Seidel for valuable discussions as well
as SCA Packaging, the German Research Foundation (DFG projects He
2789/5-1, 6-1), and the NEST program (EU project MMCOMNET) for
partial financial support.

\appendix

\section{Boundary between Damped and Growing Oscillations} \label{appendixA}
\renewcommand{\theequation}{A.\arabic{equation}}

Starting with Eq.~(\ref{EV}),
stability requires the real parts $\mbox{Re}(\lambda_i)$ of all eigenvalues $\lambda_i$ to be non-positive.
Therefore, the stability boundary is given by $\max_i \mbox{Re}(\lambda_i) = 0$. Writing
\begin{equation}
C_i + \hat{\alpha}_i D_i(1-J_{ii}) = \hat{\theta}_i + \mbox{i} \hat{\beta}_i
\end{equation}
with $C_i = P_i^0 Y_i^0 |f'_i(P_i^0)|/N_i^0$ and defining
\begin{eqnarray}
 \hat{\theta}_i &=& C_i + \hat{\alpha}_i D_i [1 - \mbox{Re}(J_{ii})] \, , \nonumber \\
 \hat{\beta}_i &=& \mp \hat{\alpha}_i D_i \mbox{Im}(J_{ii}) \mbox{ (complex conjugate eigenvalues),} \nonumber \\
 \hat{\gamma}_i &=& 4\nu_i/\mu_i{}^2 \, ,
\end{eqnarray}
we find
\begin{equation}
 2\lambda_i/\mu_i
= - \hat{\theta}_i - \mbox{i} \hat{\beta}_i + \sqrt{R_i+\mbox{i}I_i}
\label{realim}
\end{equation}
with
\begin{equation}
 R_i = \hat{\theta}_i{}^2 - \hat{\beta}_i{}^2 - \hat{\gamma}_i\hat{\theta}_i \quad \mbox{and} \quad
 I_i = 2\hat{\theta}_i\hat{\beta}_i - \hat{\gamma}_i\hat{\beta}_i \, .
\end{equation}
The real part of (\ref{realim}) can be calculated via the relation
\begin{equation}
 \mbox{Re}\Big( \sqrt{R_i \pm \mbox{i}I_i}\Big) = \sqrt{\frac{1}{2} \Big( \sqrt{R_i{}^2 + I_i{}^2} + R_i\Big) } \, .
\end{equation}
The condition $\mbox{Re}(2\lambda_i/\mu_i) = 0$ is fulfilled by $\hat{\gamma}_i = 0$ and
\begin{equation}
 \hat{\gamma}_i = 4\hat{\theta}_i (1 + \hat{\theta}_i{}^2/\hat{\beta}_i{}^2) \,  ,
\end{equation}
i.e. the stable regime is given by
\begin{equation}
 \frac{\hat{\gamma}_i}{4} = \frac{\nu_i}{\mu_i{}^2} = \frac{\hat{\alpha}_i \hat{\nu}_i}{\hat{\mu}_i{}^2} \le
 \hat{\theta}_i \left( 1 + \frac{\hat{\theta}_i{}^2}{\hat{\beta}_i{}^2} \right)
\end{equation}
for all $i$, corresponding to Eq.~(\ref{new}).\cite{mitWitt}

\section{Boundary between Damped Oscillations and Overdamped Behavior} \label{appendixB}
\renewcommand{\theequation}{B.\arabic{equation}}

For $\hat{\alpha}_i > 0$,
the imaginary parts of all eigenvalues $\lambda_i$ vanish if $\mbox{Im}(J_{ii})=0$
(i.e. $\hat{\beta}_i =0$) and if $R_i \ge 0$. This requires
\begin{equation}
\frac{4\nu_i}{\mu_i{}^2} = \hat{\gamma}_i \le \hat{\theta}_i - \frac{\hat{\beta}_i{}^2}{\hat{\theta}_i}
= \hat{\theta}_i = C_i + \hat{\alpha}_i D_i (1-J_{ii})
\end{equation}
for all $i$, corresponding to Eq.~(\ref{im}).\cite{mitWitt}


\begin{thebibliography}{000}

\bibitem{QSCM}
S. Tayur, R. Ganeshan, and M. J. Magazine,
{\it Quantitative Models for Supply Chain Management}
(Kluwer Academic, Dordrecht, 1998).

\bibitem{Zipkin} P. H. Zipkin,
{\it Foundations of Inventory Management}
(McGraw-Hill, Boston, 2000).

\bibitem{Nahmias} S. Nahmias,
{\em Production and operations analysis}
(McGraw-Hill/Irwin, Boston, 2001).

\bibitem{FactPhys}
W. J. Hopp and M. L. Spearman, {\em Factory Physics}
(McGraw-Hill, Boston, 2000).

\bibitem{mitWitt} D. Helbing, S. L\"ammer, T. Brenner, and U. Witt,
{\it Physical Review E}, in print (2004), see preprint cond-mat/0404226.

\bibitem{seba}
D. Helbing, S. L\"ammer, T. Seidel, P. \v{S}eba, and T.
P\l{}atkowski, {\it Physical Review E}, in print (2004), see
e-print cond-mat/0405230.

\bibitem{nets1}
D. J. Watts, S. H. Strogatz,
{\it Nature} \textbf{393}, 440 (1998).

\bibitem{nets2}
S. H. Strogatz,
{\it Nature} \textbf{410}, 268 (2001).

\bibitem{nets3}
R. Albert, A.-L. Barab\'{a}si,
{\it Reviews of Modern Physics}
\textbf{74}, 47 (2002).

\bibitem{nets4}
S. Maslov, K. Sneppen,
{\it Science} \textbf{296}, 910 (2002).

\bibitem{nets5}
M. Rosvall and K. Sneppen,
{\it Physical Review Letters} {\bf 91}, 178701 (2003).

\bibitem{nets6}
S. Bornholdt,  H. G. Schuster,
\textit{Handbook of Graphs and Networks}
(Wiley, Weinheim, 2003).

\bibitem{nets7}
S. Bornholdt and T. Rohlf,
{\it Physical Review Letters} {\bf 84}, 6114 
(2000).

\bibitem{Dorogovtsev}
S. N. Dorogovtsev and J. F. F. Mendes {\em Evolution of Networks}
(Oxford University Press, Oxford, 2004).

\bibitem{Dag}
C. Daganzo, {\em A Theory of Supply Chains} (Springer, New York, 2003).

\bibitem{Newell} G. F. Newell, {\em Applications of Queueing Theory}
(Chapman and Hall, Cambridge, U.K., 1982).

\bibitem{Forrester} J. W. Forrester,
{\it Industrial Dynamics}
(MIT Press, Cambridge, MA, 1961).

\bibitem{Levi} F. Chen, Z. Drezner, J. K. Ryan, and D. Simchi-Levi,
{\em Management Science} {\bf 46}(3), 436--443 (2000).

\bibitem{Cohen} M. Baganha and M. Cohen,
{\em Operations Research} {\bf 46}(3), 72-83 (1998).

\bibitem{Kahn} J. Kahn,
{\em Econom. Rev.} {\bf 77}, 667--679 (1987).

\bibitem{Lee1} H. Lee, P. Padmanabhan, and S. Whang,
{\em Sloan Management Rev.} {\bf 38}, 93--102 (1997).

\bibitem{Lee} H. L. Lee, V. Padmanabhan, and S. Whang,
{\it Management Science} {\bf 43}(4), 546 
(1997).

\bibitem{Metters} R. Metters,
in {\em Proc. 1996 MSOM Conf.}, pp. 264--269 (1996).

\bibitem{Dej02} J. Dejonckheere, S. M. Disney, M. R. Lambrecht, and D. R. Towill,
{\it International Journal of Production Economics} {\bf 78}, 133 
(2002).

\bibitem{Dej03}  J. Dejonckheere, S. M. Disney, M. R. Lambrecht, and D. R. Towill,
{\it European Journal of Operational Research} {\bf 147}, 567 
(2003).

\bibitem{Disney} S. M. Disney and D. R. Towill,
{\it International Journal of Production Research} {\bf 40}, 179 
(2002).

\bibitem{Hoberg1} K. Hoberg, U. W. Thonemann, and J. R. Bradley,
in {\em Operations Research Proceedings 2003},
Ed. D. Ahr, R. Fahrion, M. Oswald, and G. Reinelt
(Springer, Heidelberg, 2003), p. 63. 

\bibitem{Hoberg2} B. Faißt, D. Arnold, and K. Furmans,
in {\em Operations Research Proceedings 2003},
Ed. D. Ahr, R. Fahrion, M. Oswald, and G. Reinelt
(Springer, Heidelberg, 2003), p. 55. 

\bibitem{beer1}
E. Mosekilde and E. R. Larsen,
{\it System Dynamics Review}  {\bf 4}(1/2), 131 (1988).

\bibitem{beer2} J. D. Sterman,
{\em Business Dynamics}
(McGraw-Hill, Boston, 2000).

\bibitem{Arm}
D. Armbruster,
in {\em Nonlinear Dynamics of Production Systems}, Ed. G. Radons
and R. Neugebauer (Wiley, New York, 2004), p. 5.

\bibitem{Arm1}
D. Marthaler, D. Armbruster, and C. Ringhofer,
in {\em Proc. of the Int. Conf. on Modeling and Analysis of Semiconductor Manufacturing},
Ed. G. Mackulak {\em et al.} (2002), p. 365.

\bibitem{Arm2} B. Rem and D. Armbruster,
{\it Chaos} {\bf 13}, 128 (2003).

\bibitem{Arm3}
I. Diaz-Rivera, D. Armbruster, and T. Taylor,
{\it Mathematics and Operations Research} {\bf 25}, 708 (2000).

\bibitem{TGF03}
D. Helbing, in {\em Traffic and Granular Flow '03}, Ed. S.
Hoogendoorn, P. V. L. Bovy, M. Schreckenberg, and D. E. Wolf
(Springer, Berlin, 2004), see e-print cond-mat/0401469.

\bibitem{Ponzi} A. Ponzi, A. Yasutomi, and K. Kaneko,
{\it Physica A} {\bf 324}, 372 (2003).

\bibitem{Witt} U. Witt and G.-Z. Sun,
in {\em Jahrb\"ucher f. National\"okonomie u. Statistik}
(Lucius \& Lucius, Stuttgart, 2002), Vol. 222/3, p. 366.

\bibitem{NJP}
D. Helbing, {\it New Journal of Physics} {\bf 5}, 90.1--90.28 (2003).

\bibitem{inRadons}
D. Helbing, in {\em Nonlinear Dynamics of Production Systems},
Ed. G. Radons and R. Neugebauer (Wiley, New York, 2004), p. 85.

\bibitem{Radons}
G. Radons and R. Neugebauer, Eds.,  {\em Nonlinear Dynamics of Production Systems}
(Wiley, New York, 2004).

\bibitem{stabilization} T. Nagatani and D. Helbing,
{\it Physica A} {\bf 335}, 644 (2004). 

\bibitem{Nagatani}
T. Nagatani,
{\it Physica A} {\bf 334}, 243 
(2004).

\bibitem{Jorgenson}
D. W. Jorgenson,
{\em The Review of Economic Studies} {\bf 28}(2), 105--116 (1961).

\bibitem{Leontief}
W. W. Leontief,
\textit{Input-Output-Economics}
(Oxford University, New York, 1966).

\bibitem{Lefeber}
E. Lefeber, Nonlinear models for control of manufacturing systems
in {\em Nonlinear Dynamics of Production Systems}, Ed. G. Radons
and R. Neugebauer (Wiley, New York, 2004), p. 69. 

\bibitem{Eigen}
M. Eigen and P. Schuster,
{\em The Hypercycle}
(Springer, Berlin, 1979).

\bibitem{Padgett} J. Padgett,
Santa Fe Institute Working Paper No. 03-02-010.

\bibitem{Matrix} R. A. Horn and C. R. Johnson,
{\em Matrix Analysis}
(Cambridge University, Cambridge, 1985).

\bibitem{econo1}
J. B. Long jr. and C. I. Plosser,
{\it Journal of Political Economy}
\textbf{91} (1), 39 (1983).

\bibitem{econo2}
G. W. Stadler,
{\it Journal of Economic Literature}
\textbf{32} (4), 1750 (1994).

\bibitem{PRL}
L. F. Lago-Fern\'andez, R. Huerta, F. Corbacho, and J. A. Sig\"uenza, {\it Physical Review Letters}
\textbf{84}, 2758 (2000).

\bibitem{Stewart}
I. Stewart,
{\it Nature} \textbf{427}, 601 (2004).

\bibitem{disaster}
D. Helbing and C. K\"uhnert,
{\it Physica A} \textbf{328}, 584 (2003).

\bibitem{pedestrians}
D. Helbing and P. Moln\'{a}r,
{\em Physical Review E} {\bf 51}, 4282  (1995).

\bibitem{Eco}
O. N. Bj{\o}rnstad, S. M. Sait, N. C. Stenseth, D. J. Thompson,  M. Begon,
{\it Nature} \textbf{409}, 1001 (2001).

\bibitem{Dane}
S. Dane, P. G. S{\o}rensen, and F. Hynne,
{\it Nature} \textbf{402}, 320 (1999);

\bibitem{metabo1}
M. B. Elowitz and S. Leibler,
{\it Nature} \textbf{403}, 335 (2000).

\bibitem{metabo2}
E. Almaas,  B. Kov\'{a}cs, T. Vicsek, Z. N. Oltvai, and A.-L. Barab\'{a}si,
{\it Nature} \textbf{427}, 839 (2004).

\bibitem{metabo3}
A. S.  Mikhailov and B. Hess,
{\it J. Biol. Phys.} \textbf{28}, 655 (2002).

\bibitem{Ueda}
S. Koya and T. Ueda,
{\it ACH Models in Chemistry} \textbf{135} (3), 297 (1998).

\bibitem{cytrase} B. Alberts, A. Johnson,
J. Lewis, M. Raff, K. Roberts, and P. Walter, {\em Molecular
Biology of the Cell} (Garland Science, New York, 2002). 

\bibitem{calcium1} S. Schuster, M. Marhl, and T. H\"ofer,
{\em Eur. J. Biochem.} {\bf 269}, 1333 
(2002).

\bibitem{calcium2}
U. Kummer, L. F. Olsen, C. J. Dixon, A. K. Green, E. Bomberg-Bauer, and G. Baier,
{\em Biophysical Journal} {\bf 79}, 1188 
(2000).

\bibitem{calcium3}
L. Meinhold and L. Schimansky-Geier,
{\em Physical Review E} {\bf 66}, 050901(R) (2002).

\bibitem{heterotraffic} D. Helbing,
{\it Reviews of Modern Physics} {\bf 73}, 1067 (2001).

\bibitem{heteroeco}
M. Levy, H. Levy, and S. Solomon,
{\em Journal de Physique I France} {\bf 5}, 1087 
(1995).

\bibitem{Chaos} T. Beaumariage and K. Kempf,
in {\em Proceedings of the 5th IEEE/SEMI Advanced Semiconductor Manufacturing Conference},
p. 169 (1994).

\bibitem{Cha3}
T. Ushio, H. Ueda, and K. Hirai,
{\it Systems \& Control Letters} {\bf 26}, 335 (1995).

\bibitem{Cha4}
J. A. Ho{\l}yst, T. Hagel, G. Haag, and W. Weidlich,
{\it Journal of Evolutionary Economics} {\bf 6}, 31 (1996).

\bibitem{Cha5}
I. Katzorke and A. Pikovsky,
{\it Discrete Dynamics in Nature and Society} {\bf 5}, 179
(2000).

\bibitem{Peters} K. Peters, J. Worbs, U. Parlitz, and H.-P. Wiendahl,
in {\em Nonlinear Dynamics of Production Systems}, Ed. G. Radons
and R. Neugebauer (Wiley, New York, 2004), p. 39. 

\end{thebibliography}
\end{document}